\documentstyle[prd,aps]{revtex}

\begin{document}
%
%
\def\ov{\over}
\def\l{\left}
\def\r{\right}
\def\be{\begin{equation}}
\def\ee{\end{equation}}
\draft
\title{Relations between three formalisms \\ 
for irrotational binary neutron stars in general relativity} 
\author{Eric Gourgoulhon}
\address{D\'epartement d'Astrophysique Relativiste et de Cosmologie \\
  UPR 176 du C.N.R.S., Observatoire de Paris, \\
  F-92195 Meudon Cedex, France}
\date{22 April 1998}

\maketitle

\begin{abstract}
Various formalisms proposed recently for irrotational binary systems 
in general relativity are compared and explicit relations between them are
exhibited. It is notably shown that the formalisms of (i) Teukolsky, 
(ii) Shibata
and (iii) Bonazzola et al. (as corrected by Asada) are equivalent,
i.e. yield exactly the same solution, although the former two are
simpler than the latter one.  
\end{abstract}


\section{Introduction}

Inspiraling neutron star binaries are expected to be one of the
strongest source of gravitational radiation for the interferometric 
detectors GEO600, LIGO, TAMA and VIRGO currently under construction .
These systems are therefore subject to numerous theoretical studies. 
Among them are fully relativistic hydrodynamical treatments, pioneered 
by the works of Wilson et al. \cite{WilM95,WilMM96} and Oohara and 
Nakamura \cite{OohN97}. The most recent numerical calculations, those
of Baumgarte et al. \cite{BauCSST97,BauCSST98} and 
Marronetti et al. \cite{MarMW98}, rely on the
approximations of (i)
quasiequilibrium state and (ii) synchronized binaries. 
Whereas the first approximation is well justified up to the innermost stable
orbit, the second one does not correspond to physical situations, since
the gravitational-radiation driven evolution is 
too rapid for the viscous forces to synchronize the spin of each neutron star
with the orbital motion
\cite{Koc92,BilC92} --- as they do for ordinary stellar binaries. 
Rather, the fluid velocity circulation (with respect to some inertial frame)
is conserved in these systems. Provided that the initial spins
are not in the millisecond regime, this means that
close configurations are well approximated by {\em irrotational}
(i.e. with zero vorticity) states. 

The first relativistic formulation for quasiequilibrium irrotational
binaries has been given by Bonazzola, Gourgoulhon and Marck~\cite{BonGM97}
(hereafter BGM).  
Their method is based on one aspect of irrotational motion, namely the
{\em counter-rotation} (as measured in the co-orbiting frame) of the fluid
with respect to the orbital motion. Then Teukolsky~\cite{Teu98} 
and Shibata~\cite{Shi98} gave two other formulations based on the 
very definition 
of irrotationality, which implies that the specific enthalpy times the
fluid 4-velocity is the gradient of some scalar field \cite{LanL89}
({\em potential flow}). The formulation of Teukolsky is fully four-dimensional, 
whereas that of Shibata is three-dimensional. 

The aim of the present note is to clarify the relations between these
three formalisms and in particular to answer to the question: do these
rather different formalisms lead to the same solution ? Of course, at the 
Newtonian limit, they do. However it is not obvious that this  
still holds in the relativistic regime. For instance, 
the first integral of motion --- the fundamental equation which is used
to compute the matter distribution in numerical procedures --- is
written by BGM as (Eq.~(66) of Ref.~\cite{BonGM97})
\be \label{e:intprem_BGM}
    \ln h + {1\ov 2} \ln( N^2 - B_i B^i) + \ln \Gamma = {\rm const} \ , 
\ee
whereas for Teukolsky it reads (Eq.~(54) of Ref.~\cite{Teu98})
\be \label{e:intprem_T}
  N [ h^2 + D_i\psi \, D^i\psi ]^{1/2} + B^i D_i \psi = {\rm const} =: C
\ee 
and for Shibata (Eq.~(2.18) of Ref.~\cite{Shi98})
\be \label{e:intprem_S}
  {h \ov \lambda} + S^i D_i\psi  = {\rm const} \ .
\ee
In these equations, $h$ is the fluid specific enthalpy, $N$ is the lapse
function corresponding to the 3+1 slicing of spacetime by spacelike 
hypersurfaces $\Sigma_t$, $B^i$ is the
shift vector of rotating coordinates\footnote{Greek (resp. Latin) indices run 
from 0 to 3 (resp. 1 to 3). Unless explicitly mentioned, the notations
of BGM are systematically used. In particular, the 
shift vector $B^i$ is the negative of $B^i$ defined by Teukolsky \cite{Teu98}.},
$\Gamma$ is the Lorentz factor between the fluid observer and the
co-orbiting observer, $\psi$ is the scalar potential from which the fluid
4-velocity can be derived (irrotational condition), $D_i$ is the covariant
derivative on the hypersurface $\Sigma_t$ (related to the full spacetime
covariant derivative $\nabla_\alpha$ by Eq.~(33) of Ref.~\cite{Teu98}), and 
$\lambda$ and $S^i$ define the decomposition of the fluid velocity between
a part along the helicoidal Killing vector of the quasiequilibrium
assumption \cite{BonGM97} and a part parallel to $\Sigma_t$ 
\footnote{\label{ft:Shi} $\lambda$ and $S^i$ are respectively denoted
$u^0$ and $V^i$ by Shibata \cite{Shi98}.}. Needless to say,
Eqs.~(\ref{e:intprem_BGM})-(\ref{e:intprem_S}) differs from each other
quite substantially, at least at first glance. However, we shall see that
they do represent the same equation. More generally, the answer to the
question raised above is {\em yes}, i.e. BGM, Teukolsky and Shibata formalisms are
equivalent. To establish this, we will first consider Teukolsky formulation 
(Sect.~\ref{s:Teu}) and make the connection of the two other formulations
with it (Sect.~\ref{s:Shi} and \ref{s:BGM}).

\section{Teukolsky formalism} \label{s:Teu}

Let us briefly recall and discuss Teukolsky formalism \cite{Teu98}.
The neutron star matter is very well represented by a perfect fluid, whose 
stress-energy tensor writes $T_{\alpha\beta} = 
(e+p) u_\alpha u_\beta + p g_{\alpha\beta}$, $e$ being the fluid proper energy
density, $p$ the fluid pressure, $u^\alpha$ the fluid 4-velocity and 
$g_{\alpha\beta}$ the spacetime metric. The zero temperature approximation 
is fully justified for neutron star matter. In this case the fundamental 
energy-momentum conservation relation $\nabla_\mu T^{\mu\alpha} = 0$
can be shown to be equivalent to the two equations
\be \label{e:cons_imp}
	u^\mu \nabla_\mu(h u^\alpha) + \nabla^\alpha h = 0 \ , 
\ee
\be \label{e:cons_bar}
	\nabla_\mu (n u^\mu) = 0 \ , 
\ee
where $n$ is the baryon numerical density and $h$ is the specific enthalpy: 
$h=(e+p)/(m_{\rm b} n)$,  
$m_{\rm b}$ being the baryon mass. Note that thanks to the First Law of 
Thermodynamics at zero temperature, $h$ is also the baryon chemical potential
(divided by $m_{\rm b}$), so that the Gibbs-Duhem relation writes 
$\nabla_\alpha p = m_{\rm b} n \nabla_\alpha h$, relation which has been used
to derive Eq.~(\ref{e:cons_imp}).  

The {\em vorticity}, or {\em rotation 2-form}, of the fluid is defined 
as \cite{Ehl61}
\be \label{e:def_vort}
   \Omega_{\alpha\beta} = P_\alpha^{\ \,\mu} P_\beta^{\ \,\nu}
		\, \nabla_{[\nu} u_{\mu]} \ , 
\ee
where $P_{\alpha\beta} := g_{\alpha\beta} + u_\alpha\, u_\beta$ is the
projection tensor on the fluid rest-frame hyperplanes. 
Note that the definition (\ref{e:def_vort}) differs from that of Teukolsky
(Eq.~(19) in Ref.~\cite{Teu98}), which includes a factor $h$ in front of 
$u_\alpha$. However, it is easy to see that 
$\Omega_{\alpha\beta}^{\rm Teukolsky}= 2 h\, \Omega_{\alpha\beta}$, 
so that an {\em irrotational} flow can be defined either by   
$\Omega_{\alpha\beta} = 0$ or by $\Omega_{\alpha\beta}^{\rm Teukolsky}=0$.
By means of Eq.~(\ref{e:cons_imp}), the vorticity can be re-written as
\be
   \Omega_{\alpha\beta} = {1\ov h} \nabla_{[\beta} (h u_{\alpha]}) \ , 
\ee
so that the irrotational condition simply means that the 1-form $h\,u_\alpha$
is closed. Hence there exists (at least locally) a scalar field $\psi$ such that
\be \label{e:potent}
	u^\alpha = {1\ov h} \nabla^\alpha \psi \ ,  
\ee
i.e. the flow is {\em potential} \cite{LanL89}. 
The normalization relation $u_\mu u^\mu = - 1$ implies then 
\be \label{e:int_Teu_4d}
	h^2 = - \nabla_\mu \psi \, \nabla^\mu \psi \ . 
\ee
Thanks to this latter equation, the momentum conservation equation 
(\ref{e:cons_imp}) is automatically satisfied, so that the problem reduces
to finding a solution of Eq.~(\ref{e:cons_bar}), which becomes
\be \label{e:cons_bar_T_4d}
	\nabla_\mu \nabla^\mu \psi + \nabla^\mu\psi \, 
		\nabla_\mu\ln (n/h) = 0 \ .
\ee

The quasiequilibrium assumption is implemented by requiring the existence of a 
{\em helicoidal Killing vector} $l^\alpha$ \cite{BonGM97}. Teukolsky has then 
shown that Eq.~(\ref{e:int_Teu_4d}) is equivalent to Eq.~(\ref{e:intprem_T})
and that Eq.~(\ref{e:cons_bar_T_4d}) becomes (Eq.~(50) of Ref.~\cite{Teu98})
\be \label{e:eq_psi_T}
	D_i D^i \psi +  D^i\psi \, D_i\ln \l( {N n\ov h} \r)
	+ {1\ov N^2} (C - B^i D_i\psi) \l( B^i D_i \ln\l({n\ov Nh} \r) 
		- NK \r) 
	  - {1\ov N^2} B^i \, D_i ( B^j \, D_j \psi ) = 0 \ , 
\ee
where $K$ is the trace of the extrinsic curvature tensor $K_{ij}$ 
of the $\Sigma_t$ hypersurfaces and $C$ is the same constant as in 
Eq.~(\ref{e:intprem_T}). $C$ is in 
fact defined by Eq.~(39) of Ref.~\cite{Teu98}: $C = - l^\mu \nabla_\mu \psi$.
The three-dimensional equation (\ref{e:eq_psi_T}) 
has to be solved in $\psi$ to get a
solution to the problem.

\section{Links between Shibata and Teukolsky formalisms} \label{s:Shi}

The starting point of Shibata formulation \cite{Shi98} is the following
decomposition of the fluid 4-velocity (see footnote~\ref{ft:Shi}):
\be \label{e:decomp_u_Shi}
	u^\alpha = \lambda ( l^\alpha + S^\alpha) \qquad
		\mbox{with} \quad n_\mu S^\mu = 0 \ ,
\ee 
where $n^\alpha$ is the unit future-directed normal vector to $\Sigma_t$.
This relation is inserted in the momentum conservation equation
(\ref{e:cons_imp}) and the result projected onto $\Sigma_t$ to 
get\footnote{Note that 
Eqs.~(2.10) and (2.11) of Ref.~\cite{Shi98} are not correct: their 
right-hand-side must
be supplemented by respectively $-h \, n_\mu u^\mu K_{ij} V^j$ 
and $+h \, n_\mu u^\mu K_{ij} V^j$ (using Shibata's
notation for $V^j$). Fortunately, these corrections cancel each other when
summing Eqs.~(2.10) and (2.11), so that Shibata's final result (Eq.~(2.13) of
Ref.~\cite{Shi98}) is correct.} (Eq.~(2.13) of Ref.~\cite{Shi98} with 
the helicoidal symmetry taken into account)
\be \label{e:cons_imp_Shi}
   D_i \l( {h\ov \lambda} + p_i S^i \r) + S^j (D_j p_i - D_i p_j) = 0 \ , 
\ee
where
\be \label{e:def_p} 
 p^\alpha := h^\alpha_{\ \, \mu} (h u^\mu) \ , 
\ee
$h_{\alpha\beta} := g_{\alpha\beta} + n_\alpha n_\beta$ being the 
projection tensor onto $\Sigma_t$ ($p^i$ is denoted ${}^{(3)}{\tilde u}^i$
by Shibata). Shibata defines then the irrotational
condition by requiring
\be \label{e:irrot_Shi}
	p_i = D_i \phi \ , 
\ee
where $\phi$ is a scalar field. Equation~(\ref{e:cons_imp_Shi})
reduces then to the first integral (\ref{e:intprem_S}). 
By a somewhat lengthy calculation, 
Shibata shows that (\ref{e:irrot_Shi}) is equivalent to 
$\Omega_{\alpha\beta} = 0$. As discussed in Sect.~\ref{s:Teu}, this means
that the definitions of irrotationality by Shibata and Teukolsky coincide. 
It must then be possible to exhibit a correspondence between the two
formulations. 
 
First, inserting Eq.~(\ref{e:potent}) into Eq.~(\ref{e:def_p}) and using
Eq.~(\ref{e:irrot_Shi}) leads immediately to $\phi = \psi$. 
Then, the expression $l^\alpha = N n^\alpha - B^\alpha$ 
(Eq.~(7) of Ref.~\cite{BonGM97}) combined with Eq.~(\ref{e:decomp_u_Shi})
leads to 
\be \label{e:S(psi,B)}
   S^\alpha = {1\ov \lambda} h^\alpha_{\ \, \mu} u^\mu + B^\alpha 
	= {1\ov \lambda h} D^\alpha \psi + B^\alpha \ .
\ee
Besides, the normalization relation $u_\mu u^\mu = -1$ gives
\be \label{e:lamb(psi)}
   \lambda = {1\ov N} \l( 1 + {1\ov h^2} D_i\psi \, D^i \psi \r) ^{1/2} \ . 
\ee
Inserting Eqs.~(\ref{e:S(psi,B)}) and (\ref{e:lamb(psi)}) in Shibata's form
of the first integral of motion [Eq.~(\ref{e:intprem_S}) above] gives 
Teukolsky's version of it [Eq.~(\ref{e:intprem_T}) above].

The equation for determining $\psi$ given by Shibata 
(Eq.~(2.22) of Ref.~\cite{Shi98}) is also at first glance quite different
from the equation given by Teukolsky (Eq.~(\ref{e:eq_psi_T}) above), since
it reads
\be \label{e:eq_psi_S}
   D_i \l( N {n\ov h} D^i \psi \r) + D_i (N n \lambda B^i ) = 0 \ . 
\ee 
However, Eqs.~(\ref{e:lamb(psi)}) and (\ref{e:intprem_T}) yield the
following expression of $\lambda$:
\be
   \lambda = {1\ov N^2 h} ( C - B^i D_i\psi ) \ ,
\ee
which, once reported in Eq.~(\ref{e:eq_psi_S}), gives 
exactly Eq.~(\ref{e:eq_psi_T}).

\section{Links between BGM and Teukolsky formalisms} \label{s:BGM}

The BGM formulation \cite{BonGM97} is based on the fluid 3-velocity
with respect to the co-orbiting frame. 
This later frame is that of the observer whose worldlines are the trajectories
of the Killing vector $l^\alpha$, i.e. whose 4-velocity is 
$v^\alpha = e^{-\Phi}l^\alpha$, with $\Phi = 1/2 \ln(N^2 - B_i B^i)$. 
The 3-velocity $V^\alpha$ with respect to that observer is defined by the
orthogonal decomposition
\be \label{e:decomp_u_BGM}
   u^\alpha = \Gamma (V^\alpha + v^\alpha) \qquad
		\mbox{with} \quad v_\mu V^\mu = 0 \ . 
\ee
The momentum conservation equation (\ref{e:cons_imp}) is then re-expressed
as an equation for $V^\alpha$, which is a relativistic
generalization of Euler equation in a rotating frame. The basic
idea of BGM is to define a counter-rotating motion as a motion for which
the Coriolis term is canceled by the $(\nabla\wedge V)\wedge V$ term
coming from the advection term $V\cdot\nabla V$,
where the wedge product is defined in the co-orbiting observer rest frame:
$(X \wedge Y)^\alpha = v_\sigma \epsilon^{\sigma\alpha}_{\ \ \mu\nu} 
X^\mu Y^\nu$. It has been however noted by Asada \cite{Asa98} that 
the condition given by BGM is not enough to fully determine the velocity
field. In fact the correct condition is obtained by setting $F=0$ in
Eq.~(50) of BGM \cite{BonGM97}, so that this latter becomes
\be \label{e:contre_rot}
   (\nabla \wedge e^{-\Phi} V)^\alpha = - 2 e^{-\Phi} \omega^\alpha \ , 
\ee
which is identical to Eq.~(4.12) of Asada \cite{Asa98}. In this equation
$\omega^\alpha$ is the rotation vector of the co-orbiting observer:
$2\omega^\alpha := v_\sigma \epsilon^{\sigma\alpha\mu\nu}
 \nabla_\mu v_\nu$. 
If the counter-rotation condition (\ref{e:contre_rot}) is satisfied, 
the Euler equation reduces then simply to the first integral
(\ref{e:intprem_BGM}). The baryon number conservation 
equation (\ref{e:cons_bar}) gives an additional equation, for 
the divergence of $V^\alpha$:
\be \label{e:divV}
    \nabla_\mu V^\mu + V^\mu \nabla_\mu\ln(n\Gamma) = 0 \ . 
\ee
The BGM procedure is to perform a 3+1 decomposition of 
Eqs.~(\ref{e:contre_rot})-(\ref{e:divV}) and to introduce scalar and 
vector potentials so that Eq.~(\ref{e:divV}) gives a scalar 
Poisson-type equation and Eq.~(\ref{e:contre_rot}) gives a vector Poisson 
equation. This contrasts with Teukolsky or Shibata formalisms, which require
to solve only one scalar Poisson-type equation (Eq.~(\ref{e:eq_psi_T}) or
Eq.~(\ref{e:eq_psi_S})). 
It has been however noted by Asada \cite{Asa98} that the counter-rotation 
condition (\ref{e:contre_rot}) is equivalent to the irrotational 
condition $\Omega_{\alpha\beta} = 0$. Hence there must exist a 
re-interpretation of the BGM formulation in terms of Teukolsky's one. 

In fact, if the irrotational condition (\ref{e:potent}) holds, one can 
see easily, by means of the relation $\Gamma = -v_\mu u^\mu$, that
\be \label{e:gamma(C)}
	\Gamma = {C \ov h e^\Phi} \ ,
\ee
where $C$ is the same constant as in Eq.~(\ref{e:intprem_T}). 
Taking the logarithm of expression (\ref{e:gamma(C)}) gives directly the
first integral (\ref{e:intprem_BGM}).

Combining Eqs.~(\ref{e:decomp_u_BGM}), (\ref{e:potent}) and (\ref{e:gamma(C)})
yields 
\be \label{e:V(psi)}
	V^\alpha = {e^\Phi\ov C} \nabla^\alpha \psi - v^\alpha \ , 
\ee
from which one can derive the following expression for the part of $V^\alpha$
parallel to $\Sigma_t$ (cf. Sect.~V.B of BGM): 
\be
	W^i = {e^\Phi\ov C} D^i \psi + e^{-\Phi} B^i \ . 
\ee

It is immediate to verify that the 3-velocity given by Eq.~(\ref{e:V(psi)}) 
obeys to the counter-rotation condition (\ref{e:contre_rot}). 
Note that this way of proceeding is more straightforward than the converse one:
it took a full appendix in \cite{Asa98} to verify that counter-rotation
implies irrotational flow.

\section{Conclusion}

We have shown explicit relations between the formalisms introduced by
BGM, Teukolsky and Shibata to treat irrotational neutron star binaries.
All these formalisms are equivalent in the sense that they 
lead to exactly the same solution. 
Teukolsky's formulation \cite{Teu98} is the simplest one because it is 
directly based on the four-dimensional definition of irrotational motion. 
Shibata's one \cite{Shi98} contains
some complications induced by its three-dimensional character, but both
formulations result in the same type of scalar Poisson-like equation 
governing the fluid motion (Eqs.~(\ref{e:eq_psi_T}) and (\ref{e:eq_psi_S}) 
above). On the contrary, 
BGM formulation \cite{BonGM97} (corrected by Asada \cite{Asa98}) is more
complicated since it requires the resolution of an additional vector Poisson
equation. Therefore, for numerical studies, Teukolsky or Shibata procedure 
should be preferred.


\begin{references}


\bibitem{WilM95} J. Wilson and G. Mathews, Phys. Rev. Lett.
{\bf 75}, 4161 (1995).

\bibitem{WilMM96} J. Wilson, G. Mathews, P. Marronetti, 
Phys. Rev. D {\bf 54}, 1317 (1996).

\bibitem{OohN97} K. Oohara and T. Nakamura, in 
{\em Relativistic Gravitation and Gravitational Radiation},
edited by J.-A.~Marck and J.-P.~Lasota
(Cambridge University Press, Cambridge, England, 1997). 

\bibitem{BauCSST97} T.W. Baumgarte, G.B. Cook, M.A. Scheel, S.L. Shapiro,
and S.A. Teukolsky, Phys. Rev. Lett. {\bf 79}, 1182 (1997).

\bibitem{BauCSST98} T.W. Baumgarte, G.B. Cook, M.A. Scheel, S.L. Shapiro,
and S.A. Teukolsky, to be published (preprint: gr-qc/9709026).

\bibitem{MarMW98} P. Marronetti, G.J. Mathews, and J.R. Wilson,
preprint gr-qc/9803093.

\bibitem{Koc92} C.S. Kochanek, Astrophys. J. {\bf 398}, 234 (1992).

\bibitem{BilC92} L.~Bildsten and C. Cutler, Astrophys. J. {\bf 400}, 175 
(1992).

\bibitem{BonGM97}
S. Bonazzola, E. Gourgoulhon, and J.-A. Marck, Phys. Rev. D {\bf 56}, 
7740 (1997).

\bibitem{Teu98}
S.A. Teukolsky, to appear in Astrophys. J. (preprint: gr-qc/9803082).

\bibitem{Shi98}
M. Shibata, to appear in Phys. Rev. D (preprint: gr-qc/9803085).

\bibitem{LanL89}
L.D. Landau and E.M. Lifchitz, {\em M\'ecanique des fluides}, \S~134
(Editions Mir, Moscow, 1989).

\bibitem{Ehl61} J.~Ehlers, Proc. Math. Nat. Sci. Sect. Mainz Academy of
Science and Litterature {\bf 11}, 792 (1961); English translation in 
Gen. Rel. Grav. {\bf 25}, 1225 (1993).

\bibitem{Asa98}
H. Asada, to appear in Phys. Rev. D (preprint: gr-qc/9804003).




\end{references}
\end{document}